\documentclass{ws-p9-75x6-50}

\begin{document}

\title{Chaos in FRW cosmology with various types
of a scalar field potential}

\author{A. Toporensky}

\address{Sternberg Astronomical Institute, Universitetsky prospekt,13,
Moscow 119899, Russia\\E-mail: lesha@sai.msu.ru}

\maketitle

\abstracts{The results on chaos in FRW cosmology with a massive
scalar field are extended to another scalar field potential.
It is shown that for sufficiently steep potentials the chaos
disappears. A simple and rather accurate
analytical criterion for the chaos to disappear is given.
On the contrary, for gently sloping potentials the transition
to a strong chaotic regime can occur. Two examples, concerning
asymptotically flat and Damour-Mukhanov potentials are given.}

\section{Introduction}

The studies of chaotical dynamics of closed isotropic cosmological
model has a long story. They were initiated by papers \cite{Ful-Park-Star}
where the possibility to avoid a singularity at the contraction stage
in such a model
with a minimally coupled massive scalar field
was discovered. Later it was found that this model allows
the existence of periodical
trajectories \cite{Hawking} and aperiodical infinitely bouncing
trajectories having a fractal nature \cite{Page}.
 In \cite{Cornish} the
 set of  periodical trajectories was studied from the viewpoint of
 dynamical chaos theory. It was proved that the dynamics of a closed
universe with a massive scalar field is chaotic and an important
 invariant of the chaos, the topological entropy, was calculated.

However,  most of modern
scenarios based on ideas of the string theory and compactification naturally
lead to another forms of potential which are exponential or behave as
exponential for large $\varphi$ (see, for example, the paper of
Gunther and Zhuk \cite{Zhuk}).
This steepness of the
potential apparently changes the possibilities of escaping the
singularities and alters the structure of infinitely bouncing
trajectories.  Under some conditions, which are described in Section 2,
the chaotic behaviour can completely disappear.

One the other hand, potential, less steep than quadratic one can
give rise a chaotic dynamics which differ qualitatively from
described in \cite{Cornish}.  Two examples are presented in
Section 3.

\section{Chaotic properties of closed FRW model with a scalar field}
We shall consider a cosmological model with an action
\begin{equation}
S = \int d^{4} x \sqrt{-g}\left\{\frac{m_{P}^{2}}{16\pi} R +
\frac{1}{2} g^{\mu\nu}\partial_{\mu}\varphi \partial_{\nu}\varphi
-V(\varphi)\right\}.
\end{equation}
For a closed Friedmann model with the metric
\begin{equation}
ds^{2} = dt^{2} - a^{2}(t) d^{2} \Omega^{(3)},
\end{equation}
where
$a(t)$ is a cosmological radius,
$d^{2} \Omega^{(3)}$ is the metric of a unit 3-sphere and
with homogeneous scalar field $\varphi$
we have the following ODE system:
\begin{equation}
\frac{m_{P}^{2}}{16 \pi}\left(\ddot{a} + \frac{\dot{a}^{2}}{2 a}
+ \frac{1}{2 a} \right)
+\frac{a \dot{\varphi}^{2}}{8}
-\frac{a V(\varphi)}{4} = 0,
\end{equation}
\begin{equation}
\ddot{\varphi} + \frac{3 \dot{\varphi} \dot{a}}{a}
+ V'(\varphi) = 0.
\end{equation}
where the scalar field potential $V(\phi)$ is a smooth nonnegative function
with $V(0)=0$.

This system has one first integral of motion
\begin{equation}
-\frac{3}{8 \pi} m_{P}^{2} (\dot{a}^{2} + 1)
+\frac{a^{2}}{2}\left(\dot{\varphi}^{2} + 2 V(\varphi)\right)  =
0.
\end{equation}

It is easy to see from Eq. (5) that the points of maximal expansion
and those of minimal contraction, i.e. the points, where $\dot{a} =
0$ can exist only in the region where
\begin{equation}
a^{2} \leq \frac{3} {8 \pi}  \frac{m_{P}^2}{V(\varphi)} ,
\end{equation}
Sometimes, the region defined by inequalities (6) is called Euclidean,
and the opposite region is called
Lorentzian. We will use this definition for brevity, though we consider
these notations rather meaningless \cite{Quant}.

It also can be shown that
the possible points of
maximal expansion are localized inside the region
\begin{equation}
a^{2} \leq \frac{1}{4 \pi} \frac{m_{P}^{2}}{V(\varphi)}
\end{equation}
while the possible points of minimal contraction lie outside this
region (7) being at the same time inside the Euclidean region
(6) \cite{our2}.

In our paper \cite{our} it was shown that the region of
possible points of maximal expansion has quite a regular structure.
In the left, closest to axis $a = 0$ part of this region there are
points of maximal expansion after which trajectory goes to
singularity.
Then one can see the region where after the going
through the point of maximal expansion a trajectory undergoes the
``bounce'' i.e. goes through the point of minimal contraction. Then
we have the region where after going through the point of maximal
expansion trajectory has a ``$\varphi$ - turn '' i.e. has the
extremum in the value of the scalar field $\varphi$ and then falls
into a singularity. Then one has the region corresponding to the
trajectories having bounce after one oscillation in $\varphi$ and
so on.

To avoid a misunderstanding, let us indicate once more what initial
condition space we use. When we start from maximal expansion point,
we fix one time derivative ($\dot a=0$). So, for fix the initial
condition completely, we need only to specify initial values of
$a$ and $\varphi$. The initial $\dot \varphi$ is determined from
the constraint equation (5), and our initial condition space becomes
($a$, $\varphi$). We will study the structure of this space,
investigating the location of points of maximal expansion, starting
from which a trajectory has a bounce, but not the location of
bounce itself.  The $\varphi=0$ cross-section
of regions, leading to bounce, are called as "intervals".

Remembering that a trajectory
describing an expanding universe must have a point of maximal expansion,
we can further apply this analysis to bouncing trajectories. Their second
point of maximal expansion may lie either inside "bounce" regions,
defined at the first step or between them. This fact generates a more tiny
substructure of the region under consideration. This
substrucure of regions having two bounces repeats the
general structure of regions having at least one bounce
and so on and so forth. Continuing this process
{\it ad infinitum} we can get the fractal zero-measure
set of infinitely bouncing trajectories escaping the singularity.

Numerical investigations show also
that all  simple periodical trajectories (i.e. having only one bounce
per period) have a full stop point on the Euclidean boundary (the
curve defined by the equality in (6)). Moreover,
trajectories, going from the boundary inside the Euclidean region
has a point of maximal expansion
almost immediately
and then go towards
a singularity. So, periodical trajectory approaches their bounce point
on the boundary from the Lorentzian side. Hence, we have a necessary
condition for
given region of the Euclidean boundary to contain a full stop points
of periodical trajectories (it is the condition that a trajectory starting
with zero velocities from the Euclidean boundary goes into the Lorentzian
region):
 \begin{equation}
\frac{\ddot{\varphi}}{\ddot{a}} < \frac{d \varphi}{da},
  \end{equation}
where function $\varphi(a)$ in the right-hand side is the equation
of the Euclidean boundary (6).

The point on the Euclidean boundary where a trajectory starting
with zero velocities have a direction tangent to the Euclidean boundary
can be obtained substituting the equality
in (6) into the equations of motion.
It was first introduced by Page \cite{Page} for massive scalar field
potential $V(\phi)=m^2 \phi^2/2$. In this case

 \begin{eqnarray}
&& \varphi_{page} = \sqrt{\frac{3}{4\pi}} m_{P}; \nonumber\\
&&a_{page}=1/m ,
\end{eqnarray}
 except for the
trivial solution $\varphi=0$, $a= \infty$.

The first periodical trajectory have full stop point at
$\varphi=1.19 \varphi_{page}$ and $a=0.83 a_{page}$.
These obtained numerically values, rather close to the Page point,
represent a right-down boundary of a set of full stop points of
periodical trajectories on the Euclidean boundary.

Using the equation of motion (3)-(4) the criterion (8) gives for an
arbitrary scalar field potential
  \begin{equation}
 V(\varphi)>\sqrt{\frac{3
m_{P}^2}{16 \pi}}V'(\varphi)
 \end{equation}

The condition (10) may be treated as restricting a local steepness of
the function $V(\phi)$. In can be easily seen that for power-law
function the Page value $\phi_{page}$ such that for all $\phi >
\phi_{page}$ the condition (8) is satisfied, always exists. For steeper
potentials the situation changes. For example, the potential
$V(\varphi)=M_{0}^4(\cosh(\varphi/\varphi_0)-1)$, studied in \cite{our2}
has a Page point only if
$\varphi_0 > \frac{\sqrt{3}}{4 \sqrt{\pi}} m_P$. In the opposite case
the condition (8) is never satisfied and all the trajectories
starting from the Euclidean boundary go into the Euclidean region
and soon experience the point of maximal expansion.  It was
confirmed numerically that the chaos is absent in this case
(see Fig.1.).

\begin{figure}
\epsfxsize=20pc 

\epsfbox{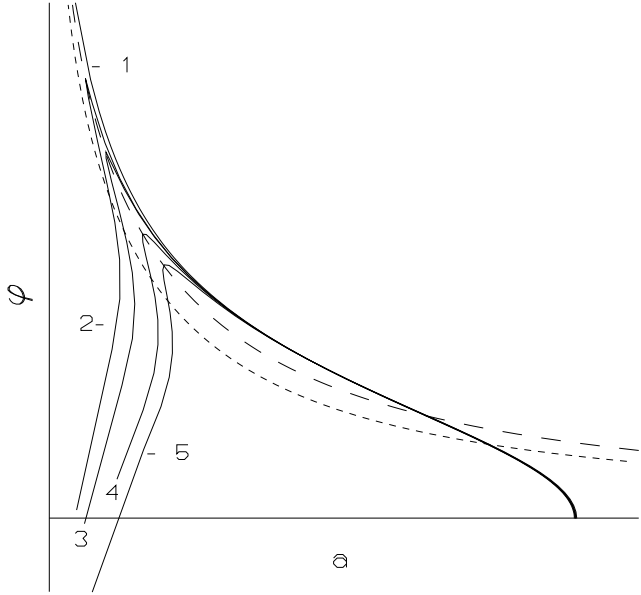} 
\caption{Example of trajectories with the initial conditions
close to
the boundary separating trajectories falling
into $\varphi=+ \infty$ (trajectory $1$) and $\varphi=-\infty$
(trajectories $2-5$) singularities for the case $\varphi_0 <
\frac{\sqrt{3}}{4 \sqrt{\pi}} m_P$. This boundary is sharp, no fractal
structure is present.  Trajectories $2-5$ have a zigzag-like form,
no periodical trajectories are present. The long-dashed line is the
Euclidean boundary, the short-dashed line is the separating curve.}
\end{figure}

For steeper potentials like
$$
 V(\varphi)=
M_{0}^4 (\exp(\varphi^2/\varphi_{0}^2)+\exp(-\varphi^2/\varphi_{0}^2)-2)
$$
inequality (8) is definitely violated for large $\phi$ but, depending on $\phi_0$
it can be satisfied for intermediate $\phi$.
 For different $\phi_0$
there are two or zero
Page points in this case. The value of $\phi_0$ corresponding to the
Page points disappearance ($0.905 m_P$) differ from the $\phi_0$
of the chaos disappearence obtained numerically ($0.96 m_P$)
for $\sim 6\%$ \cite{our2}. It is interesting that when the chaos
exists, the number of
bounce intervals is finite and the whole picture of the chaos is
similar to the picture obtained for a system with a massive scalar field
 and a hydrodynamical matter
\cite{ournew}.

 So, studying the possibility to satisfy the condition (8) at the
Euclidean boundary,
we have an easily calculated and rather accurate
criterion for the existence of the chaotic dynamics in the system (3)-(5).

\section{Chaotic dynamics for gently sloping potentials}

In this section we  describe the opposite case - the potential
which is less steep than the quadratic one. We will see that in this case the
transition to a qualitatively stronger chaos may occur.

\subsection {Asymptotically flat potentials and merging of the bounce
intervals}

Let us start with the potential
\begin{equation}
V(\varphi)=M_{0}^{4}(1-exp(-\frac{\varphi^2}{\varphi_{0}^{2}})),
\end{equation}
where $M_0$ and $\varphi_0$ are parameters. $M_0$ determines the asymptotical
value of the potential for $\varphi \to \pm \infty$.

It can be easily checked from the equations of motion that multiplying
the potential to a constant (i.e. changing the $M_0$) leads only to
rescaling $a$. So, this procedure do not change the
chaotic properties of our dynamical system. On the contrary,
this system appear
to be very sensitive to the value of $\varphi_0$. We plotted in Fig.2.
the $\varphi=0$ cross-section of bounce intervals depending on $\varphi_0$.
In our numerical investigations we use the units in which
$m_P/\sqrt{16 \pi}=1$, and below we present our results in these units
because now most of the interesting events
occur for the range of parameters of the order of unity.
This plot represents a situation, qualitatively different from studied
previously for potentials like $V \sim \varphi^2$ and steeper. Namely,
the bounce intervals can merge.

\begin{figure}
\epsfxsize=20pc 

\epsfbox{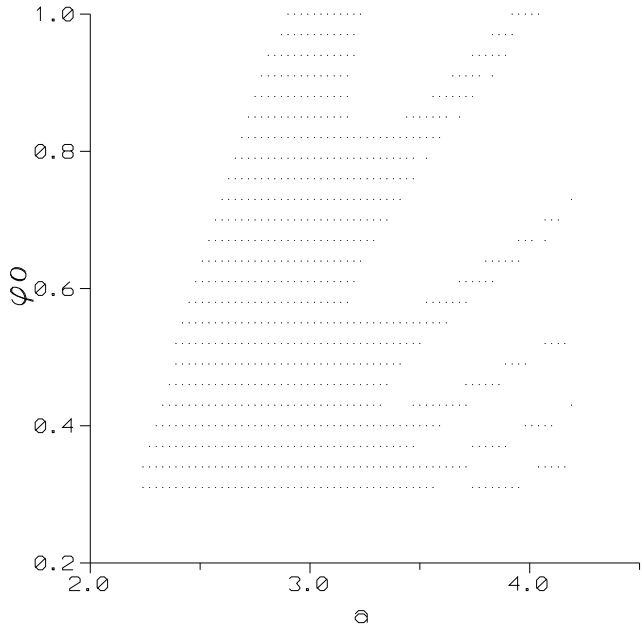} 
\caption{
The $\varphi=0$ cross-section of the bounce intervals for the potential
(11) depending on $\varphi_0$. Consecutive merging of $5$ first intervals
can be seen in this range of $\varphi_0$.
}
\end{figure}

Let us see more precisely what does it means. For $\varphi_0>0.82$ the picture
is qualitatively as for a massive scalar field - trajectories from 1-st
interval have a bounce with no $\varphi$-turns before it,
trajectories which have initial point of maximal expansion between
1-st and 2-nd intervals fall into a singularity after one $\varphi$-turn,
 those from
2-nd interval have a bounce after one $\varphi$-turn and so on. For $\varphi_0$
a bit smaller than the first merging value the 2-nd interval contains
trajectories with two $\varphi$-turns before bounce, the space between
1-st interval (which is now the product of two merged intervals)
and the 2-nd one contains trajectories falling into a singularity
after two $\varphi$-turns. There are no trajectories going to a singularity
with exactly one $\varphi$-turn.
Trajectories from the 1-st
interval can experience now
a complicated chaotic behavior which can not
be described in as similar way as above.

With $\varphi_0$ decreasing further, the process of interval merging
being to continue leading to growing chaotisation of trajectories.
When $n$ intervals merged together, only trajectories with at
least $n$ oscillations of the scalar field before falling into
a singularity are possible. Those having exactly $n$ $\varphi$-turns
have their initial point of maximal expansion between 1-st bounce interval
and the 2-nd one (the second interval now contains trajectories having a bounce after
$n$ $\varphi$-turns). For initial values of the scale factor larger then
those
from the 2-nd interval, the regular
quasiperiodic structure described above is restored.

 Numerical analysis shows also
that the fraction of very chaotic trajectories as a function of
$\varphi_0$ grows rapidly with $\varphi_0$ decreasing below the first
merging value. To illustrate this point we
studied the behavior of trajectories starting from the point of maximal
expansion with initial values of $a$ located in the range of the first
two intervals on the greed  with step $0.002$. Total number of trajectories
for each $\varphi_0$ is equal to $1000$. We
 plotted in Fig.3 the number of
trajectories which do not fall into a singularity during first $50$
oscillations of the scalar field $\varphi$.  We do not include
trajectories with the next point of maximal expansion located outside
the 2-nd (or the 1-st one, if merging occurred) interval, so all
counted trajectories avoid a singularity during this sufficiently long
 time interval due to their extreme chaoticity, but not due to reaching
the slow-roll regime.
 Before merging, the measure of so chaotic trajectories is
extremely low and they are undistinguishable on our grid. When $\varphi_0$
becomes slightly lower than the value of the first merging, this number
begin to grow rather rapidly and for $\varphi_0 \sim 0.6$ near $10 \%$ of
trajectories from the 1-st interval on our grid experience at least 50 oscillation
before falling into a singularity.

\begin{figure}
\epsfxsize=20pc 

\epsfbox{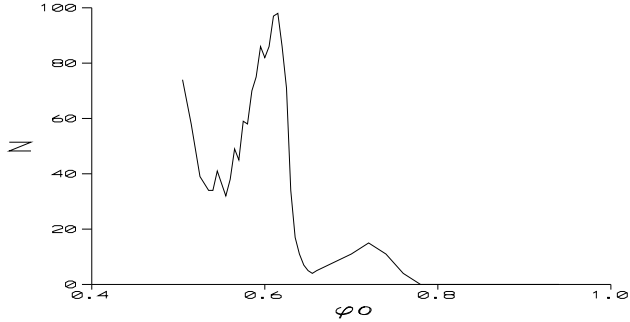} 
\caption{
Number $N$ of trajectories do not falling into a singularity during
$50$ oscillating times for the potential (11) depending on the parameter
$\varphi_0$. The scale factor of the initial maximal expansion point varies
in the range of the 1-st and 2-nd intervals which merge at $\varphi_0=0.82$.
Total number of trajectories is equal to $1000$.
}
\end{figure}

We point out that for a simple massive scalar field potential only
$\sim 10^{-2}$ trajectories in the same range of the initial scale factors
have at least one bounce (using the abovementioned
quasiperiodicity in the structure of initial condition space
($a$, $\varphi$) we can estimate this measure as a ratio of interval
width and distance between near-by intervals). The width of subintervals
containing
trajectories not falling into a singularity after only one bounce
is about one hundred times less than the width of "main" intervals
and so on. The common numerical
calculation accuracy is unsufficient for distinguishing even the sole
trajectory with 50 oscillations and $a$ being in the range of first two
intervals.

In contrast to this, the chaos
for the potential (11) is really significant. Detail of intervals merging
including the description out of $\varphi=0$ cross-section require further
analysis.

For large  initial $a$ the configuration of bounce intervals
for potential (11) looks like the configuration for
a massive scalar field potential with the effective
mass $m_{eff}=(\sqrt{2} M_{0}^{2})/\varphi_0$.
The periods of corresponding structures coincides with a good accuracy
($m \Delta a \sim 2.8$ in the initial condition space)
though the widths of the intervals for the potential (11) is
bigger then for $V=(m_{eff}^2 \varphi^2)/2$.

\subsection{Damour-Mukhanov potentials}
The very chaotic regime described above is possible also for potentials,
which are not asymptotically flat, if the potential growth is slow enough.
We will illustrate this point describing
  a particular (but rather wide) family of
potentials having power-low behavior -- Damour - Mukhanov potentials
\cite{Damour}.
They was originally introduced to show a possibility to have an
inflation behaviour without slow-roll regime. After, various issues on
inflationary dynamics \cite{Liddle}
and growth of perturbation \cite{Taruya,Cardenas}
for this kind of scalar
field potential was studied.

The explicit form of Damour-Mukhanov potential is
\begin{equation}
V(\varphi)=\frac{M_{0}^{4}}{q} \left[ \left(1+\frac{\varphi^2}
{\varphi_{0}^{2}} \right)^{q/2}-1 \right],
\end{equation}
with three parameters --$M_0$, $q$ and $\varphi_0$.

For $\varphi \ll \varphi_0$ the potential looks like the massive one with the
effective mass $m_{eff}=M_{0}^{2}/\varphi_0$. In the opposite case of large
$\varphi$ it grows like $\varphi^q$.

As in the previous section, the chaotic behavior does not depend on
$M_0$.  So, we have a two-parameter family of potentials with different
chaotic properties. Numerical studies with respect to possibility of
bounce intervals merging shows the following picture (see Fig.4): for
a rather wide range of $q$ there exists a corresponding critical value
of $\varphi_0$ such that for $\varphi_0$ less than critical, the very
chaotic regime exists. Increasing $q$ corresponds to decreasing
critical $\varphi_0$.

\begin{figure}
\epsfxsize=20pc 

\epsfbox{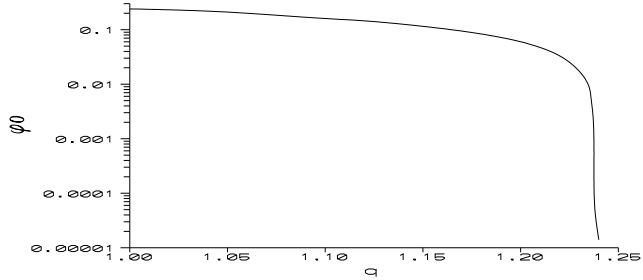} 
\caption{
The value $\varphi_0$ of the potential (12) corresponding to the
first merging of the bounce intervals depending on $q$.
}
\end{figure}

 Surely, since this regime is absent for quadratic and more
steep potentials, $q$ must at least be less than $2$. We can see clearly
the very chaotic regime for $q< 1.24$.
 The case $q=1.24$ lead to strong chaos for $\varphi_0<1.4 \times 10^{-5}$ and
the critical $\varphi_0$ decreases with increasing $q$ very sharply at this
point. We did not investigated further these extremely small values of
$\varphi_0$, because the physical meaning of such kind of potential is
very doubtful.

\end{document}